\documentclass[twocolumn]{aastex631} 
\usepackage[fleqn]{amsmath}
\usepackage{soul}

\newcommand\jpz[1]{\textcolor[rgb]{1,0,0}{{\scriptsize JP:} #1}}

\shorttitle{Saturn's Moons Colors}
\shortauthors{Pe\~na et al.}

\graphicspath{{./}{figures/}}

\begin{document}

\title{Colors of Irregular Satellites of Saturn with DECam}

\correspondingauthor{Jos\'e Pe\~na}
\email{jpena@das.uchile.com}

\author[0000-0002-4552-4743]{Jos\'e Pe\~na}
\affiliation{Departamento de Astronom\'ia, Universidad de Chile, Camino del Observatorio 1515, Las Condes, Santiago, Chile.}
\affiliation{Centro de Excelencia en Astrofísica y Tecnologías Afines (CATA), Chile}
\affiliation{Millennium Institute of Astrophysics (MAS), Chile}

\author{Cesar Fuentes}
\affiliation{Departamento de Astronom\'ia, Universidad de Chile, Camino del Observatorio 1515, Las Condes, Santiago, Chile.}

\begin{abstract}

We report $g-r$ and $r-i$ new colors for 21 Saturn Irregular Satellites, among them, 4 previously unreported.
This is the highest number of Saturn Irregular satellites reported in a single survey. These satellites were measured by ``stacking'' their observations to increase their signal without trailing.
This work describes a  novel processing algorithm that enables the detection of faint sources under significant background noise and in front of a severely crowded field.

Our survey shows these new color measurements of Saturn Irregular Satellites are consistent with other Irregular Satellites populations as found in previous works and reinforcing the observation that the lack of ultra red objects among the irregular satellites is a real feature that separates them from the trans--Neptunian objects (their posited source population). 
 
\end{abstract}


\section{Introduction} \label{sec:intro}

Irregular satellites (Irrs) are minor bodies characterized by large orbits with high inclinations and/or high eccentricities, many of them retrograde, in contrast with the nearly circular, co-planar and compact orbits of regular satellites. They have been found orbiting the giant planets and because of their extreme orbits it is assumed that this is a population captured at some point during the Solar System evolution \citep{1956VA......2.1631K}.

Recent modeling of the dynamical evolution of the Solar System (SS) has built a consensus towards giant planets changing their orbits since they first assembled \citep{2005Natur.435..459T} and evidence of this migration has been found in the distribution of several different populations of small bodies: Main Belt \citep{1997AJ....114..396G, 2009Natur.457.1109M, 2010AJ....140.1391M}, Jovian Trojans and Neptunian Trojans (JTs and NTs respectively) \citep{2000Icar..148..479F, 2005Natur.435..462M, 2010MNRAS.405.1375L, 2016A&A...592A.146G},  and Trans-Neptunian Objects \citep{2008Icar..196..258L}.
In the case of Irrs, simulations run by \cite{2007AJ....133.1962N, 2014ApJ...784...22N} showed that they could have been captured from nearby bodies in the planetesimal disk during planet migration in similar numbers as those observed, and consistent with similarities in their size distributions \citep{2006AJ....132..171S}.

Trans--Neptunian Objects (TNOs) are the least evolved remnants of the planetesimal disk, in part because a relatively lower solar irradiation that has not affected their surface properties as other closer population. 
Since IRRs are expected to be captured TNOs during the early history of the giant planet dynamical evolution, it is expected that the color properties of TNOs and Irrs should be similar.
It has been found that TNOs show a high diversity of sizes and colors, from red to ultra red bodies \citep{2010AJ....139.1394S, 2012AJ....144..169S, 2015A&A...577A..35P, 2017AJ....154..101P, 2018PASJ...70S..40T, 2018PASJ...70S..38C, 2019AJ....158...53T}. At the same time,  colors of many Irrs have been measured (around half of their known populations), especially those from Jupiter (JIrrs) and Saturn (SIrrs) \citep{2000Icar..143..371S, 2001Icar..154..313R, 2003Icar..166...33G, 2004ApJ...605L.141G, 2004ApJ...613L..77G, 2006Icar..184..181B, 2007Icar..191..267G, 2015ApJ...809....3G, 2018AJ....155..184G, 2018RNAAS...2...42M}, finding that there are no ultra-red bodies among the Irrs as among the TNOs, raising questions about their actual origin and their evolution. Hence the necessity of gathering more data about these bodies to have a more complete perspective and see if current models are still valid or new ones are necessary.

In this work we present observations pointing at known SIrrs aiming to detect sources until magnitude $\sim R=25$ (namely, almost all SIrrs known until then, July, 2019). To detect such faint objects is necessary to use the ``pencil-beam'' or ``shift and stack'' method \citep{1998AJ....116.2042G, 2004Natur.430..865H, 2004Icar..169..474K, 2009ApJ...696...91F}. We observed in $gri$ in order to obtain $g-r$ and $r-i$ colors. 
Despite observing against crowded fields and having non-photometric nights, we report the highest number of SIrrs colors obtained in one single survey so far (21 SIrrs with $r-i$ and 16 with $g-r$). In section \ref{sec:data} we explain our observing plan and the resulting data; in section \ref{sec:ImAnalysis} we explain the data processing we used to obtain SIrrs photometry; in section \ref{sec:results} we show our color results for the detected SIrrs; in section \ref{sec:discussion} we compare our results with colors from other works and discuss about their sources of error; in section \ref{sec:conclusions} we summarize our main findings and put them in the context of other populations and current SS evolution models, as well as laying out future work and lessons for future surveys that aim at obtaining pencil-beam photometry.


\section{Data} 
\subsection{Survey Strategy} \label{sec:data}
Our observations were made with the Dark Energy Camera (DECam) mounted at the prime focus of the Blanco 4 meters telescope at the Cerro-Tololo International Observatory (CTIO\footnote{\url{https://noirlab.edu/public/programs/ctio/}}). DECam covers a 3 square degree field of view with a mosaic of $\sim60$ ccd of 2Kx4K pixels, yielding a  $0.27\arcsec$/pixel resolution \citep{2008SPIE.7014E..0ED}. These characteristics allowed us to cover almost all SIrrs using only two different pointings.

We observed the region surrounding Saturn during 4 nights between the 2nd and the 5th of July, 2019. We used two different pointings every night to account for Saturn's motion 
We refer to these two areas \emph{``Field 1''} and \emph{``Field 2''}. 
These fields were designed to image as many satellites as possible while keeping Saturn off the field. These change every night to account for parallax and targets falling in between chips.
The bright planet's proximity imposed important background gradients in our images. Additionally, Saturn was in the vicinity of the galactic plane, increasing the chance of our targets to be observed near field stars. 
Our careful data reduction was fairly successful in accounting for these features (see section \ref{sec:ImProc}) 

Most SIrrs are too faint to be detected in a single exposure without trailing, so we considered many short exposures to be ``integrated'' later, accounting for the target's sky motion (more details of this ``stacking'' process in section \ref{sec:photcoadd}).
The fields were imaged in $g$, $r$ and $i$ bands with 120 second exposures to prevent trailing. A handful of short exposures (15 and 30 seconds) in $g$, $r$, $i$ and $z$ bands where also taken every night. In Table \ref{tab:survey} we detail our observations.

Since we wanted to detect irregular satellites brighter than $r\sim25$ on each filter each night, and taking into account that they are known to be red bodies, we designed our survey to observe them $\gtrsim$1 hour in $r$ and $i$ and $\gtrsim$2 hours in $g$. As seen in Table \ref{tab:survey}, we divided our observations in single filter blocks. Both fields were observed in $r$ and $i$, but constrained by the night's duration, we could observed only one field in $g$ (causing that each field was observed only 2 nights in $g$). Short observations were designed to take $gri$ bands in between blocks.

\begin{deluxetable*}{C|c c c|C|C|C|c}
\tablecaption{Survey Chronology
\label{tab:survey}}
\tablecolumns{8}
\tablewidth{0pt}
\tablehead{
\colhead{Night\tablenotemark{a}} & \colhead{Field\tablenotemark{b}} & \colhead{R.A.\tablenotemark{c}} & \colhead{Decl.\tablenotemark{c}}  & \colhead{Filter\tablenotemark{d}} &  \colhead{Exp. Time\tablenotemark{e}} & \colhead{N\tablenotemark{f}} & \colhead{Start\tablenotemark{g}} 
}
\startdata
1 & 1 & 19h11m13s & 287d48m14s & griz\tablenotemark{*} & 15 & 1 & 2019-07-03 01:09:46 \\
 &  &  &  & r & 120 & 35 & 2019-07-03 02:41:01 \\
 &  &  &  & i & 120 & 34 & 2019-07-03 04:09:36 \\
 & 2 & 19h19m26s & 289d51m27s & griz\tablenotemark{*} & 15 & 1 & 2019-07-03 04:12:05 \\
 &  &  &  & i & 120 & 35 & 2019-07-03 05:39:58 \\
 &  &  &  & r & 120 & 35 & 2019-07-03 07:08:39 \\
 & 1 & 19h11m13s & 287d48m14s & griz\tablenotemark{*} & 15 & 1 & 2019-07-03 07:11:07 \\
 &  &  &  & g & 120 & 60 & 2019-07-03 09:41:44 \\
 &  &  &  & griz\tablenotemark{*} & 15 & 1 & 2019-07-03 09:44:13 \\
 &  &  &  & g & 120 & 10 & 2019-07-03 10:09:20 \\
\hline
2 & 2 & 19h19m09s & 289d47m20s & griz\tablenotemark{*} & 30 & 1 & 2019-07-04 01:01:28 \\
 &  &  &  & r & 120 & 35 & 2019-07-04 02:32:40 \\
 &  &  &  & i & 120 & 35 & 2019-07-04 04:03:15 \\
 & 1 & 19h10m52s & 287d42m54s & griz\tablenotemark{*} & 30 & 1 & 2019-07-04 04:05:45 \\
 &  &  &  & r & 120 & 35 & 2019-07-04 05:39:53 \\
 &  &  &  & i & 120 & 32 & 2019-07-04 07:09:05 \\
 & 2 & 19h19m09s & 289d47m20s & griz\tablenotemark{*} & 30 & 1 & 2019-07-04 07:11:33 \\
 &  &  &  & g & 120 & 30 & 2019-07-04 08:30:48 \\
 &  &  &  & griz\tablenotemark{*} & 30 & 1 & 2019-07-04 08:33:16 \\
 &  &  &  & g & 120 & 30 & 2019-07-04 09:53:00 \\
\hline
3 & 1 & 19h09m51s & 287d27m52s & griz\tablenotemark{*} & 30 & 1 & 2019-07-04 09:55:29 \\
 &  &  &  & r & 120 & 35 & 2019-07-05 02:35:19 \\
 &  &  &  & i & 120 & 35 & 2019-07-05 04:04:59 \\
 & 2 & 19h18m56s & 289d44m04s & griz\tablenotemark{*} & 30 & 1 & 2019-07-05 04:08:48 \\
 &  &  &  & r & 120 & 35 & 2019-07-05 05:42:44 \\
 &  &  &  & i & 120 & 34 & 2019-07-05 07:13:36 \\
 & 1 & 19h09m51s & 287d27m52s & griz\tablenotemark{*} & 30 & 1 & 2019-07-05 07:16:04 \\
 &  &  &  & g & 120 & 53 & 2019-07-05 09:34:08 \\
\hline
4 & 2 & 19h18m37s & 289d39m15s & griz\tablenotemark{*} & 30 & 1 & 2019-07-05 09:36:36 \\
 &  &  &  & r & 120 & 30 & 2019-07-06 02:20:30 \\
 &  &  &  & i & 120 & 30 & 2019-07-06 03:41:06 \\
 & 1 & 19h09m31s & 287d22m49s & griz\tablenotemark{*} & 30 & 1 & 2019-07-06 03:43:33 \\
 &  &  &  & r & 120 & 30 & 2019-07-06 05:11:08 \\
 &  &  &  & i & 120 & 30 & 2019-07-06 06:26:54 \\
 & 2 & 19h18m37s & 289d39m15s & gri\tablenotemark{*} & 30 & 1 & 2019-07-06 06:29:22 \\
 &  &  &  & g & 120 & 60 & 2019-07-06 09:05:11 \\
 &  &  &  & gri\tablenotemark{*} & 30 & 1 & 2019-07-06 09:07:40 \\
 &  &  &  & g & 120 & 11 & 2019-07-06 09:37:31 \\
\enddata
\tablenotetext{a}{Number of the night.}
\tablenotetext{b}{Number of the Field (1 or 2).}
\tablenotetext{c}{Central coordinates of the field observed (note that they change each night).}
\tablenotetext{d}{Filter used in the observations.}
\tablenotetext{e}{Duration of a single exposure (in seconds).}
\tablenotetext{f}{Number of observations taken on the specified filter.}
\tablenotetext{g}{Time when the observations started (in UTC).}
\tablenotetext{*}{When exposures are short (15 or 30 seconds), one exposure on each filter was taken consecutively.}
\end{deluxetable*}

\subsection{Satellites' Localization} 
To locate known SIrrs we used the ephemeris provided by the Jet Propulsion Laboratory (JPL) web service\footnote{JPL Horizons: \url{https://ssd.jpl.nasa.gov/horizons/}}, obtained via \texttt{astroquery}\footnote{\url{https://astroquery.readthedocs.io/en/latest/}} \citep{2019AJ....157...98G}, which is a coordinated package of \texttt{astropy}\footnote{\url{https://www.astropy.org/}} \citep{2013A&A...558A..33A, 2018AJ....156..123A}, both packages built in \texttt{Python}\footnote{\url{https://www.python.org/}}. We observed that the bodies' phase angle, the heliocentric distance and the geocentric distance barely change through the four nights, as shown in Table \ref{tab:satranges}.

Because of their brightness, we were able to detect Hyperion and Phoebe in the short exposure images (Hyperion and Iapetus were the two regular satellites visible, but the latter was saturated). For all other SIrss, ``stacked'' magnitudes were measured (see section \ref{sec:photcoadd}).

\begin{deluxetable}{rcrrr}
\tablecaption{Observation Geometry
\label{tab:satranges}}
\tablecolumns{5}
\tablewidth{0pt}
\tablehead{
 & \colhead{Param.\tablenotemark{a}} & \colhead{mean} & \colhead{min.} & \colhead{max.}
}
\startdata
All\tablenotemark{b} & $\alpha$ & 0.527 & 0.232 & 0.818 \\
 & $r$ & 10.067 & 9.889 & 10.271 \\
 & $\Delta$ & 9.054 & 8.874 & 9.261 \\
\hline
Saturn\tablenotemark{c} & $\alpha$ & 0.529 & 0.352 & 0.698 \\
 & $r$ & 10.050 & 10.050 & 10.050 \\
 & $\Delta$ & 9.037 & 9.035 & 9.040 \\
\hline
Satellites\tablenotemark{d} & $\alpha$ & & 0.339 & 0.351 \\
 & $r$ & & 0.000 & 0.010 \\
 & $\Delta$ & & 0.001 & 0.014 \\
\enddata
\tablenotetext{a}{Orbital Parameters: Phase (Sun-Object-Earth) angle ($\alpha$, in degrees), heliocentric distance ($r$ in au) and geocentric distance ($\Delta$, in au).}
\tablenotetext{b}{Values for all bodies (Saturn and its satellites) at observed times.}
\tablenotetext{c}{Values for Saturn at observed times.}
\tablenotetext{d}{Minimum and maximum variation of each parameter for each satellite at observed times.}
\end{deluxetable}

\section{Image Analysis} \label{sec:ImAnalysis}

\subsection{Image Processing} \label{sec:ImProc}
All image processing was done using the methods implemented in \texttt{photutils}\footnote{\url{https://photutils.readthedocs.io/en/stable/}} \citep{larry_bradley_2020_4049061}, which is an affiliated package of \texttt{astropy}. First, we computed the images' two-dimensional ``background'' and its root mean square (called for now on ``background-RMS'' or just ``RMS'')\footnote{Using \texttt{Photutils}' function \texttt{Background2D}}. Both ``background'' and ``background-RMS'' are calculated and used in a similar way to \texttt{SExtractor}\footnote{\url{https://www.astromatic.net/software/sextractor/} \\ \url{https://sextractor.readthedocs.io/}} \citep{1996A&AS..117..393B}, separating the image in a grid of cells, estimating the mode\footnote{The mode estimator is equal to $2.5\mathrm{median}-1.5\mathrm{mean}$ (or the median if $(\mathrm{mean} - \mathrm{median}) /\mathrm{std} > 0.3$) where the median, mean and standard deviation (std) are calculated over the ``sigma-clipped'' pixels (rejecting values farther than 3 standard deviations from the median) for each cell in the grid.} (as the background) and the standard deviation (as the root mean square) on each cell to form a ``low resolution map'' of both ``background'' and ``background-rms''. A median filter is performed in both ``low resolution maps'' to finally interpolate them to the size of the original image (using spline interpolation of order 3). 
Using visual inspection, we explored the resulting backgrounds trying different sizes for cells (30-300 pixels) and median filters (1-11 pixels). We chose a cell size of 62x62 pixels, which is small enough to distinguish between the background from the many stars that occupy these crowded fields and to account for the diversity of background gradients and patterns caused by the proximity of Saturn. Yet, this cell size tends to overestimate the background near bright stars. To compensate for that effect we set the size of the median filter to 9 pixels for the ``low resolution map'' (instead of 3, which is the size the algorithm recommends by default). This led us to underestimate the fluxes of bright stars ($r\leq15.5$) but did not affect the result for most sources.
Once the background and the background-rms were calculated, we were ready to do photometry to the static sources in the background-subtracted images, using the uncertainties obtained from the background and the background-rms as weight maps (as explained in section \ref{sec:phot}).


Prior to doing photometry on SIrrs (which are transient sources with the risk of falling on top or nearby stars) we visually inspect around their expected positions and check for contamination by nearby sources.
We constructed a ``Sky'' image for each CCD with all transient objects, like asteroids and satellites, removed.
This ``Sky'' with only static sources is then subtracted from each image, allowing us to further clean our images from most of the contribution of sources nearby the satellites.
We begun by taking the background-subtracted images and moved them to a common frame (corresponding to the image closest to the mean coordinate of those taken in the same night and the same band) using ``spline interpolation''. We considered rotation and translation that best matched representative positions between their WCS\footnote{``World Coordinate System'', transforming from pixel values to ``sky'' coordinates (namely, right ascension $\alpha$ and declination $\delta$) or vice versa. For this, we used \texttt{astropy.wcs}.} as computed by the DECam repository. 
We produced the Sky as the median (in each pixel) of all interpolated images. Simply using the images median works fine for all satellites but the brightest, for which a trail of their trajectory is left. Using Phoebe as reference (the SIrr that leaves the brightest trail) we realize that excluding the 20 pixels closest to the Irr on each image was enough to get rid of that trail. Another way of doing the same was implementing a ``sigma-clipping'' process through each pixel to reject outlier values (when a satellite or other moving object is crossing that pixel). We selected which technique worked best for each satellite by visually inspecting the final result. The satellites that left a trail in the Sky image and required this processing were Phoebe, Albiorix, Siarnaq, Paaliaq, Ymir and Kiviuq. Finally, we simply subtract the Sky from each single exposure image.

\subsection{Photometry} \label{sec:phot}
We performed forced photometry on known sources. 
This photometry consists in fitting a two-dimensional gaussian profile on the stamp of the source (a small piece of the background subtracted image centered at the source). This gaussian profile is defined in equation \ref{eq:gauss}, where $f(x,y)$ is the fitted flux at the pixel coordinate $(x,y)$ of the stamp and the total flux is $2\pi F_0 \sigma_x \sigma_y$.
\begin{equation}
    f(x,y) = F_0 \exp\left(-\frac{x-x_0}{2\sigma_x^2} - \frac{y-y_0}{2\sigma_y^2}\right)
    \label{eq:gauss}
\end{equation} 

We used the inverse square of an ``error-map'' in that stamp as the weights of the fitting process.
This error-map, similarly to the one computed by \texttt{SExtractor}, accounts for the background error (the RMS) and the Poisson error in each pixel\footnote{Implemented with the \texttt{calc\_total\_error} function of \texttt{photutils}, where the error in each pixes is equal to $\sqrt{\sigma^2_{bkg} + F/g_{eff}}$, with $\sigma^2_{bkg}$ and $F$ the RMS and the flux in that pixel and $g_{eff}$ the effective gain (namely, the number of electrons per flux count).}. 
To fit these Gaussian profiles to the sources we used the Levenberg-Marquardt algorithm\footnote{Implemented with the \texttt{LevMarLSQFitter} method of \texttt{astropy}.} \citep{Levenberg.1944, Marquardt.1963}. Considering that our images have FWHM between 3 and 9 pixels, as reported by the DECam repository, we tested our results using stamps of 19x19 and 11x11 pixels, giving both similarly good results (reflected by our errors when calculating the magnitudes, as explained in section \ref{sec:getmags}). Since most of our fields are relatively crowded, we chose 11x11pix stamps to reduce contamination from nearby sources.

First, to detect the sources we used \texttt{SExtractor} on the original images, configured to calculate the background and its RMS as described in section \ref{sec:ImProc} and to detect sources with brightness  of al least twice the RMS.
We chose \texttt{SExtractor} due to its ability to deblend sources and to judge their probability of being stars\footnote{We used the old \texttt{SExtractor}'s neural network classifier which gives a probability of being a star (in contrast of being a galaxy). For this it requires the seeing of the image, which we provided using the FWHM reported in the DECam repository images.}. \texttt{SExtractor} is able to do all of this rapidly and with low computational cost. To fit the profiles of these sources, we only used those that are farther than 150 pixels away from CCD edges (avoiding ``empty'' pixels left over from shifting images to a common frame, as explained at the end of section \ref{sec:ImProc}), that have positive fluxes and a probability of being stars  higher than 0.7, as computed by \texttt{SExtractor} (since we found that sources that did not satisfy that requirement were extended, blended or spurious). 

Having fit all sources, we calculated the ``sigma'' that characterized the width of the profile of all sources ($\sigma$ from equation \ref{eq:gauss}). To do that, we select only sources that are at least 11 pixels away from the closest source, then we reject all sources too bright or faint (clipping at 2-sigma in brightness) 
to then reject all sources with a standard deviation in the fitted residual (the stamp minus the fitted profile) larger than twice the median RMS in that stamp. Finally, with the remaining sources, we calculate the ``sigma'' of the profile as the mean between $\overline{\sigma_x}$ and $\overline{\sigma_y}$ (the median of the individual profile ``sigmas'' in the CCD's $x$ and $y$ coordinates respectively, after performing a final ``sigma-clipping'' over the profile's ``sigma''). All the selections and rejections above were aimed to reject any artefact, blended, saturated, and poorly fit source.
Once we had the characteristic ``sigma'' across each CCD, we re-fit sources with a fixed point spread function to get the fluxes we report in this work.
This is done independently for each epoch and CCD, delivering all the information necessary to transform the SIrrs fluxes to magnitudes (see section \ref{sec:getmags}).

It is worthy to mention that we tried several methods before settling on the one above. We started by using the aperture photometry from \texttt{SExtractor}, but the high background variability created very noisy detections. Then we used the \texttt{photutils} version of \texttt{DAOPHOT} \citep{1987PASP...99..191S}, but its method to fit overlapping sources simultaneously produced very inaccurate results and took a long time to execute.
Finally we implemented the algorithm explained above, which uses a simple PSF model that does not need to detect sources (since we only use known sources) nor fit several sources simultaneously (since we focus in isolated bodies).


\subsection{Stacked Images} \label{sec:photcoadd}
To determine the colors of Saturn satellites we must ``stack'' our images (namely, integrate a series of images to increase the signal-to-noise ration of the target source until it is detectable, as in \citealt{1998AJ....116.2042G, 2004Natur.430..865H, 2004Icar..169..474K, 2009ApJ...696...91F}) to account for most of them being too faint to be detected in single exposure images.
We considered 11x11 sky-subtracted stamps centered on the expected satellites' positions to stack them.
For each night and band a different stack was produced. 
The main difference between our method and the one presented in \cite{1998AJ....116.2042G} is that we do not scale the fluxes of the images to equalize them to a reference, not to create the Sky image nor for the stack. Instead, we visually inspect each single exposure stamp to ensure only stamps not contaminated by bright stars halos or other features, such as cosmic rays or detector artifacts, are included in the stack. The advantages of not scaling fluxes are that we do not add uncertainty to the observations (such as from imperfect scaling factors or PSF-matching) and we do not increase the noise contribution of certain images, in particular those that suffer more extinction and are usually noisier.

On each stacked stamp we measure the source's flux as explained in section \ref{sec:phot}. We fit a two-dimensional Gaussian profile, only this time we weight the pixels using the inverse square of the standard deviation calculated in the vicinity of the source. 
To fit the Gaussian profile of the satellite we previously needed the ``sigma'' of the profile of a stacked source, so we also stacked all images that contribute to the satellite's stacked stamp but without removing the Sky to finally get that ``sigma'' from the stacked stars, using the same technique explained in section \ref{sec:phot} (namely, fitting individually the ``sigma'' of the profiles of the sources detected using \texttt{SExtractor} and getting the general ``sigma'' as the median of the individual values after removing outliers).
Finally, we rejected measurements with low ($<$3) signal-to-noise, calculated as the amplitude of the fitted Gaussian profile divided by the root mean square of the residual stamp. (the stamp minus the fitted profile).

\subsection{Obtaining Magnitudes} \label{sec:getmags}

Once all fluxes were obtained (see section \ref{sec:phot}), we compute their transformation to magnitudes. We considered the PanSTARRS photometry as reference \citep{2016arXiv161205560C}. We got all stars from PanSTARRS on the CCDs that contain our target satellites\footnote{Obtained via \texttt{astroquery.mast}, querying for Data Release 2 values} and matched them to our sources (using their sky coordinates, right ascension $\alpha$ and declination $\delta$). We only considered stars closer than $1\arcsec$ to each source (since the cross-match distances were mostly on that range). PanSTARRS provide ``Mean'' and ``Stack'' photometries (computed over single exposure images and stacked images respectively), having ``PSF'' and ``Aperture'' magnitudes in both cases. We used the ``Mean PSF'' photometry (and astrometry) since it showed more proximity with our data \citep[consistent with the photometry and astrometry performance shown in][]{2016arXiv161205560C}. To improve our photometry 
we used a similar subset of stars as the ones used to obtained the ``sigma'' of our sources' profile, namely, stars farther than 150 pixels from the CCD borders and at $5\arcsec$ (around 20 pixels) farther from the closest source. Then, we also rejected sources with PanSTARRS magnitudes (``$\ast$\texttt{MeanPSFMag}'') brighter than 16 (which were typically saturated), and magnitude uncertainties (``$\ast$\texttt{MeanPSFMagErr}'') and standard deviations (``$\ast$\texttt{MeanPSFMagStd}'') above 0.05 and 0.1 respectively.

To transform our fluxes to magnitudes we used equation \ref{eq:phot}, where $n$ and $t$ are the identifiers of the source and the time they were observed. For each CCD and time, we calculate $m^0 = 2.5\log_{10}(Q)$, where $Q$ is the ``sigma-clipped'' mean of $F_{n,t}/F_{n}$, with $F_{n,t}$ the measured flux of the source $n$ at time $t$ and $F_{n}$ the flux of that source according to PanSTARRS (namely, $F_{n}=10^{-0.4m_{PS1,n}}$, with $m_{PS1}$ equal to ``$\ast$\texttt{MeanPSFMag}''). This gives one $m^0$ for all sources in one CCD at one time. Having $m_{n,t}$, $F_{n,t}$ and $m_{n,t}^0$ known for all selected PanSTARRS sources (with $m_{n,t}=m_{PS1,n}$), we obtained the parameters $A$ by solving equation \ref{eq:phot} in matrix form including all values of known PanSTARRS sources (where $(g-r)_n$, $(r-i)_n$ and $(r-i)_n$ are the colors of the source $n$ and $X_{n,t}$ is equal to the airmass proxy $\sec(z_{n,t})$ with $z$ the zenith angle of the source at time $t$). Here $A_3$, $A_4$ and $A_5$ account for color corrections, $A_1$ for the airmass extinction and $A_2$ accounts for a general extinction that affects all sources at a given time. All $A$ parameters are calculated (and used) separately for each night, band and field.

\begin{equation}
    \begin{split}
    m_{n,t} =& -2.5\log_{10}(F_{n,t}) + m_{n,t}^0 + X_{n,t} A_1 + A_{2,t} \\ 
    &+ A_3(g-r)_n + A_4(g-i)_n + A_5(r-i)_n
    \label{eq:phot}
    \end{split}
\end{equation}

Given the nature of this work, we are interested in doing photometry in stacked sources. Since a stacked flux $\langle F \rangle$ is computed on the mean of a series of stamps ($i=1,...,N$), we have that $\langle F \rangle=\frac{1}{N}\sum F_i$, where $F_i = 10^{-0.4(m-f_i)}$ (obtained by expressing equation \ref{eq:phot} as $m=-2.5\log_{10} F_i + f_i$), we obtained equation \ref{eq:mag_stack}, which give us the magnitude corresponding to the stacked source (using all the $m^0_i$ and $A$ parameters obtained when solving equation \ref{eq:phot} to calculate $f_i$).

\begin{equation} \label{eq:mag_stack}
    \begin{split}
    \langle m \rangle= -2.5\log_{10}\langle F \rangle + \langle m \rangle_0 \\
    \langle m \rangle_0 = 2.5\log_{10} \left( \frac{\sum_i^N 10^{0.4f_i}}{N} \right)
    \end{split}
\end{equation}

For each exposure $i$ we consider two sources of error, one from the measured flux and other from the standard deviation of the errors of the known sources when solving equation \ref{eq:phot}. The later is taken as the error of $f_i$ (namely, $\sigma_{f_i}$). Propagating those errors to obtain $\sigma_{\langle m \rangle_0}$ (the error of $\langle m \rangle_0$) using a Taylor's expansion, we obtained equation \ref{eq:stack_err}. However, this underestimates $\sigma_{\langle m \rangle_0}$ for each SIrr so we report the root mean square of the errors on the stacked known sources instead.

\begin{equation} \label{eq:stack_err}
    \sigma_{\langle m \rangle_0}^2 = 
    \frac{\sum_i^N \left(\sigma_{f_i}10^{0.4f_i} \right)}{\left(\sum_i^N 10^{0.4f_i} \right)^2}
\end{equation}

To get $f_i$ in each case we need to know the source's colors, which are unknown at the beginning. We use an iterative procedure assuming no color correction at first and then using the resulting magnitudes in each band to start correcting the colors until the magnitudes vary less than $\sigma_{f_i}/2$.

\section{Results} \label{sec:results}
From the short exposure images we measured Hyperion (the only regular satellite observed that was not saturated) and Phoebe (the brightest SIrr). Since these observations visited one field using each band consecutively, we derived their magnitudes assuming one set of colors per visit (remember that, as explained in section \ref{sec:getmags} we needed to refine our magnitudes by iterating) 
The result is shown in Table \ref{tab:colors_short}. Note that there are observations in $z$ band, since we also included color terms relative to that band when solving equation \ref{eq:phot}.

\begin{deluxetable*}{l|cc|C|C C C}
\tablecaption{Magnitudes and Colors of Hyperion and Phoebe
\label{tab:colors_short}}
\tablecolumns{7}
\tablewidth{0pt}
\tablehead{
\colhead{Name} & \colhead{Night} & \colhead{UTC\tablenotemark{a}} &
\colhead{$r$} & \colhead{$g-r$} & 
\colhead{$r-i$} & \colhead{$i-z$}
}
\startdata
Hyperion & 2 & 2019-07-04 01:04:17.760 & 14.25 \pm 0.03 & 0.51 \pm 0.05 & 0.23 \pm 0.04 & 0.01 \pm 0.05 \\
 &  & 2019-07-04 04:07:41.664 & 14.32 \pm 0.03 & 0.48 \pm 0.05 & 0.17 \pm 0.04 & 0.05 \pm 0.05 \\
 &  & 2019-07-04 07:13:29.856 & 14.26 \pm 0.03 & 0.51 \pm 0.05 & 0.16 \pm 0.04 & 0.08 \pm 0.05 \\
 &  & 2019-07-04 08:35:11.328 & 14.26 \pm 0.03 & 0.53 \pm 0.05 & 0.16 \pm 0.04 & 0.04 \pm 0.05 \\
 &  & 2019-07-04 09:57:28.224 & 14.28 \pm 0.03 & 0.42 \pm 0.05 & 0.15 \pm 0.04 & 0.16 \pm 0.05 \\
 & 4 & 2019-07-06 03:45:31.104 & 14.07 \pm 0.03 & 0.47 \pm 0.04 & 0.22 \pm 0.04 & 0.07 \pm 0.04 \\
\hline
Phoebe & 1 & 2019-07-03 04:13:30.720 & 16.52 \pm 0.03 & 0.34 \pm 0.04 & 0.13 \pm 0.04 & -0.02 \pm 0.04 \\
 &  & 2019-07-03 09:45:38.880 & 16.39 \pm 0.03 & 0.36 \pm 0.04 & 0.14 \pm 0.04 & -0.06 \pm 0.04 \\
 & 3 & 2019-07-05 01:05:13.056 & 16.47 \pm 0.02 & 0.37 \pm 0.03 & 0.11 \pm 0.03 & -0.02 \pm 0.03 \\
 &  & 2019-07-05 04:10:48.288 & 16.49 \pm 0.02 & 0.34 \pm 0.03 & 0.14 \pm 0.03 & -0.04 \pm 0.03 \\
 &  & 2019-07-05 07:19:31.008 & 16.34 \pm 0.02 & 0.33 \pm 0.03 & 0.14 \pm 0.03 & -0.04 \pm 0.03 \\
 &  & 2019-07-05 09:38:33.792 & 16.44 \pm 0.02 & 0.36 \pm 0.03 & 0.11 \pm 0.03 & -0.01 \pm 0.03 \\
\enddata
\tablenotetext{a}{Start of  the $r$ band observation.}
\end{deluxetable*}

\begin{figure}
\centering
\includegraphics[width=\hsize]{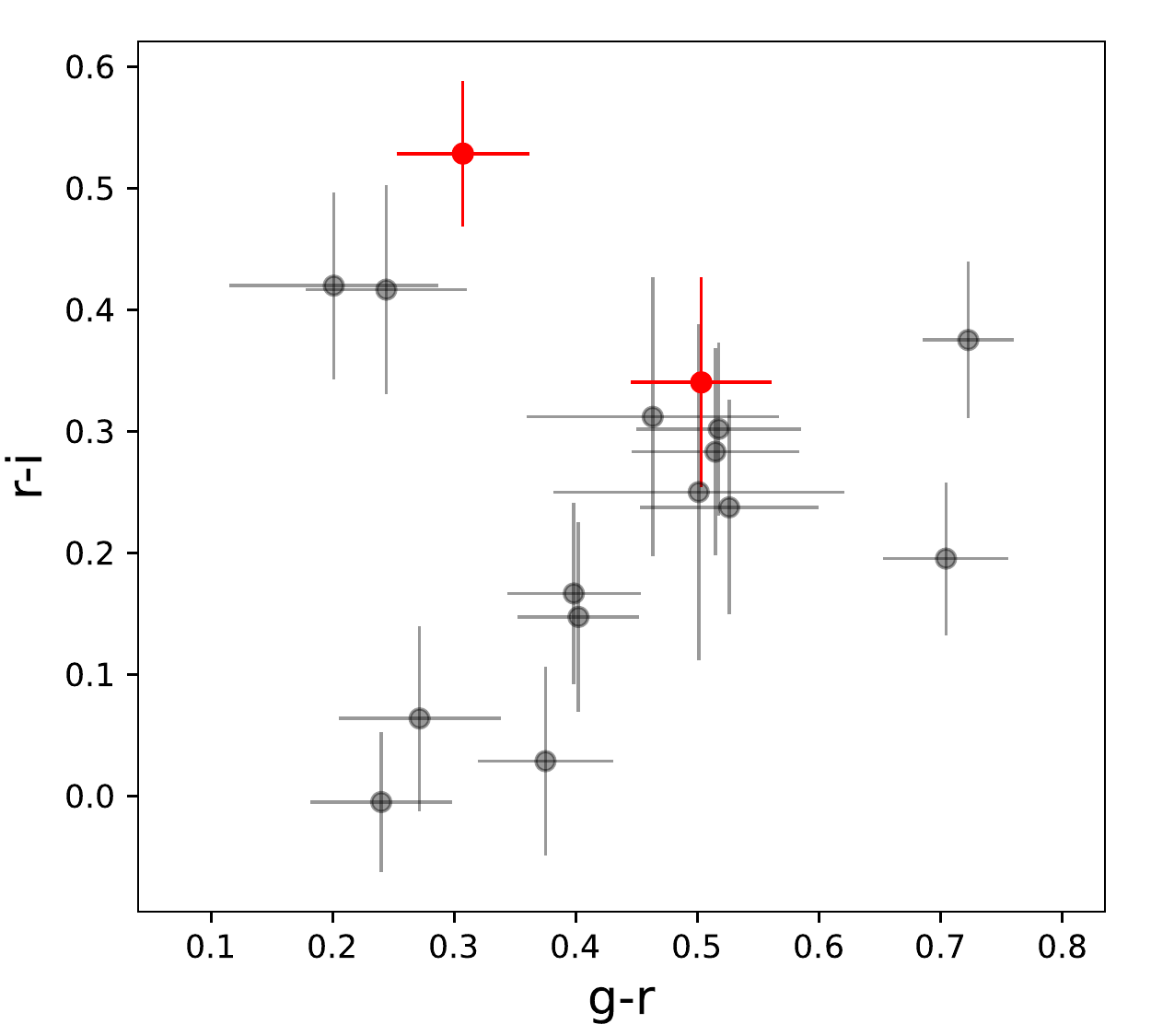}
\caption{Color-color plot for values from table \ref{tab:colors_stack}. In red are Hyrrokkin and Kari, bodies whose colors had never been measure before (note that Skoll and Loge had not been measured before either, but they were measure  in $r$ and $i$ only). The error bars were obtained by propagating the individual errors when computing the mean magnitudes and the colors.}
\label{fig:colors}
\end{figure}

We also obtained magnitudes and colors from stacked sources. 
In the case of stacked images, to obtain an equivalent to the mean magnitude through the different SIrr's rotational phases, we calculate the average between stacked measurements. The colors reported in table \ref{tab:colors_stack} (and used when solving equation \ref{eq:phot}) were obtained using those mean magnitudes.
We plot these colors in Figure \ref{fig:colors}, with Hyrrokkin and Kari in red, since their colors are reported for the first time in this work.
In Appendix \ref{appendix:colors} we list all magnitudes used to compute colors.

\begin{deluxetable}{l|C|C C}
\tablecaption{Satellites magnitudes and colors
\label{tab:colors_stack}}
\tablecolumns{4}
\tablewidth{0pt}
\tablehead{
\colhead{Name} & \colhead{$r$\tablenotemark{a}} & 
\colhead{$g-r$\tablenotemark{b}} & 
\colhead{$r-i$\tablenotemark{b}}
}
\startdata
Phoebe & 16.49 \pm 0.06 & 0.27 \pm 0.07 & 0.06 \pm 0.08 \\
Siarnaq & 20.47 \pm 0.05 & 0.53 \pm 0.07 & 0.24 \pm 0.09 \\
Albiorix & 21.07 \pm 0.09 & 0.46 \pm 0.10 & 0.31 \pm 0.11 \\
Ymir & 22.25 \pm 0.11 & 0.50 \pm 0.12 & 0.25 \pm 0.14 \\
Tarvos & 22.88 \pm 0.05 & 0.52 \pm 0.07 & 0.30 \pm 0.07 \\
Kiviuq & 22.49 \pm 0.06 & 0.51 \pm 0.07 & 0.28 \pm 0.09 \\
Paaliaq & 21.51 \pm 0.04 &  & 0.25 \pm 0.05 \\
Ijiraq & 23.18 \pm 0.05 & 0.70 \pm 0.05 & 0.20 \pm 0.06 \\
Skathi & 24.26 \pm 0.04 & 0.24 \pm 0.06 & -0.005 \pm 0.093 \\
Bebhionn & 24.66 \pm 0.05 &  & 0.24 \pm 0.09 \\
Erriapus & 23.69 \pm 0.05 & 0.20 \pm 0.09 & 0.42 \pm 0.08 \\
Skoll & 25.02 \pm 0.05 &  & 0.36 \pm 0.08 \\
Hyrrokkin & 24.08 \pm 0.06 & 0.50 \pm 0.08 & 0.34 \pm 0.09 \\
Mundilfari & 24.36 \pm 0.05 & 0.38 \pm 0.06 & 0.03 \pm 0.08 \\
Narvi & 24.48 \pm 0.06 & 0.24 \pm 0.07 & 0.42 \pm 0.09 \\
Suttungr & 24.32 \pm 0.06 & 0.40 \pm 0.07 & 0.17 \pm 0.08 \\
Bestla & 24.64 \pm 0.04 & 0.72 \pm 0.07 & 0.38 \pm 0.07 \\
Thrymr & 24.08 \pm 0.05 & 0.40 \pm 0.06 & 0.15 \pm 0.08 \\
Kari & 24.65 \pm 0.06 & 0.31 \pm 0.07 & 0.53 \pm 0.09 \\
Loge & 24.80 \pm 0.05 &  & 0.15 \pm 0.08 \\
Fornjot & 24.93 \pm 0.06 &  & 0.20 \pm 0.09 \\
\enddata
\tablenotetext{a}{Magnitude mean observed each night.}
\tablenotetext{b}{Colors obtained from the mean magnitudes observed each night.}
\end{deluxetable}

\section{Discussion} \label{sec:discussion}

\begin{figure}
\centering
\includegraphics[width=\hsize]{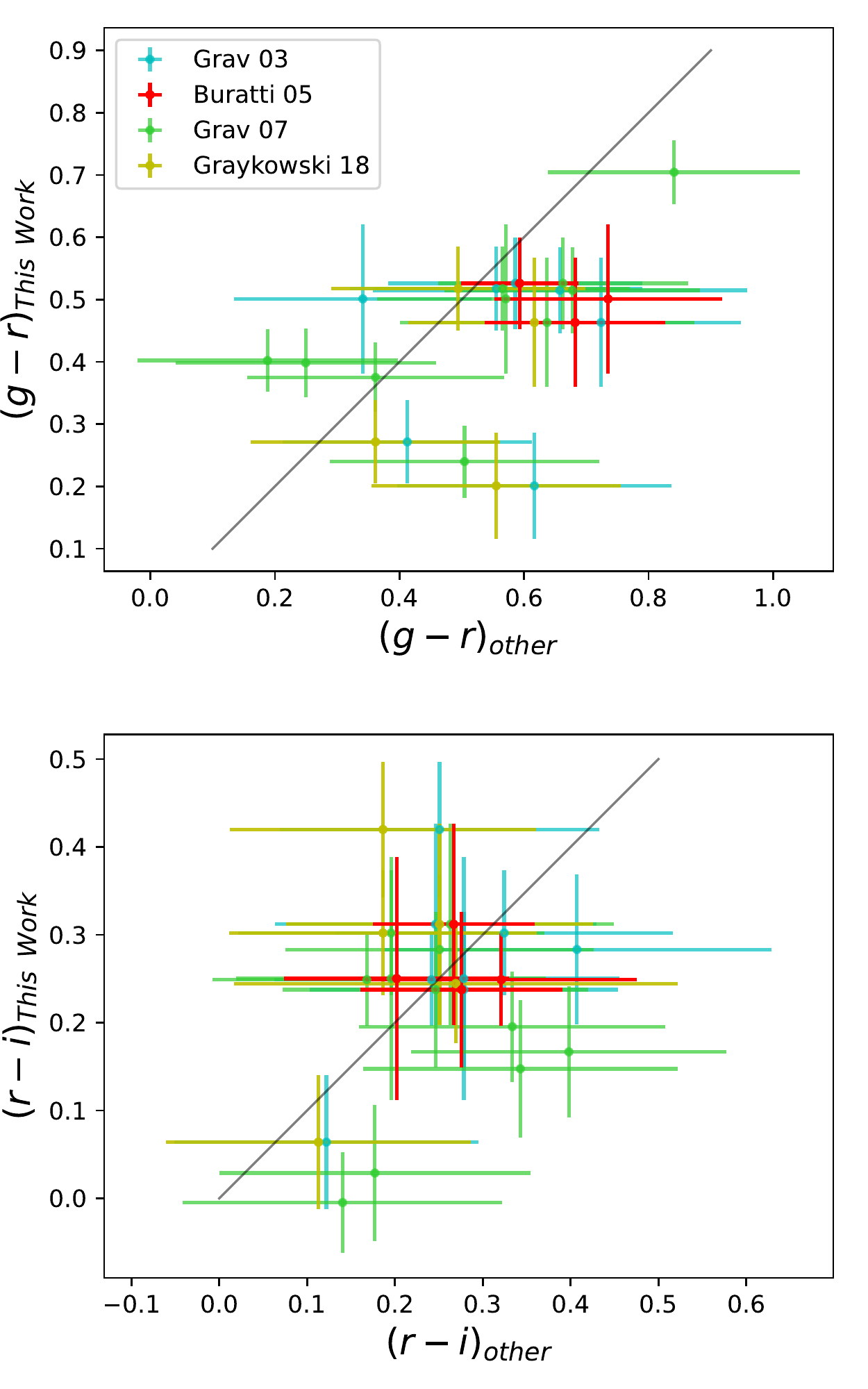}
\caption{Our colors compared to colors from other works. In light blue, colors from \cite{2003Icar..166...33G}, in red from \cite{2005Icar..175..490B}, in green from \cite{2007Icar..191..267G} and in yellow from \cite{2018AJ....155..184G}. In black is the one-to-one line.}
\label{fig:color_comparison}
\end{figure}

We compared our measurements with those from other works. The result is shown in Figure \ref{fig:color_comparison}. Notice that all results are reported in the $BVR$ bands, with the exception of  \cite{2005Icar..175..490B} which does so in $gri$ filters. To transform $BVR$ to $gri$ we used the equations from \cite{2005AJ....130..873J}. We observe a high dispersion between our results and those of other works.

\begin{figure}
\centering
\includegraphics[width=\hsize]{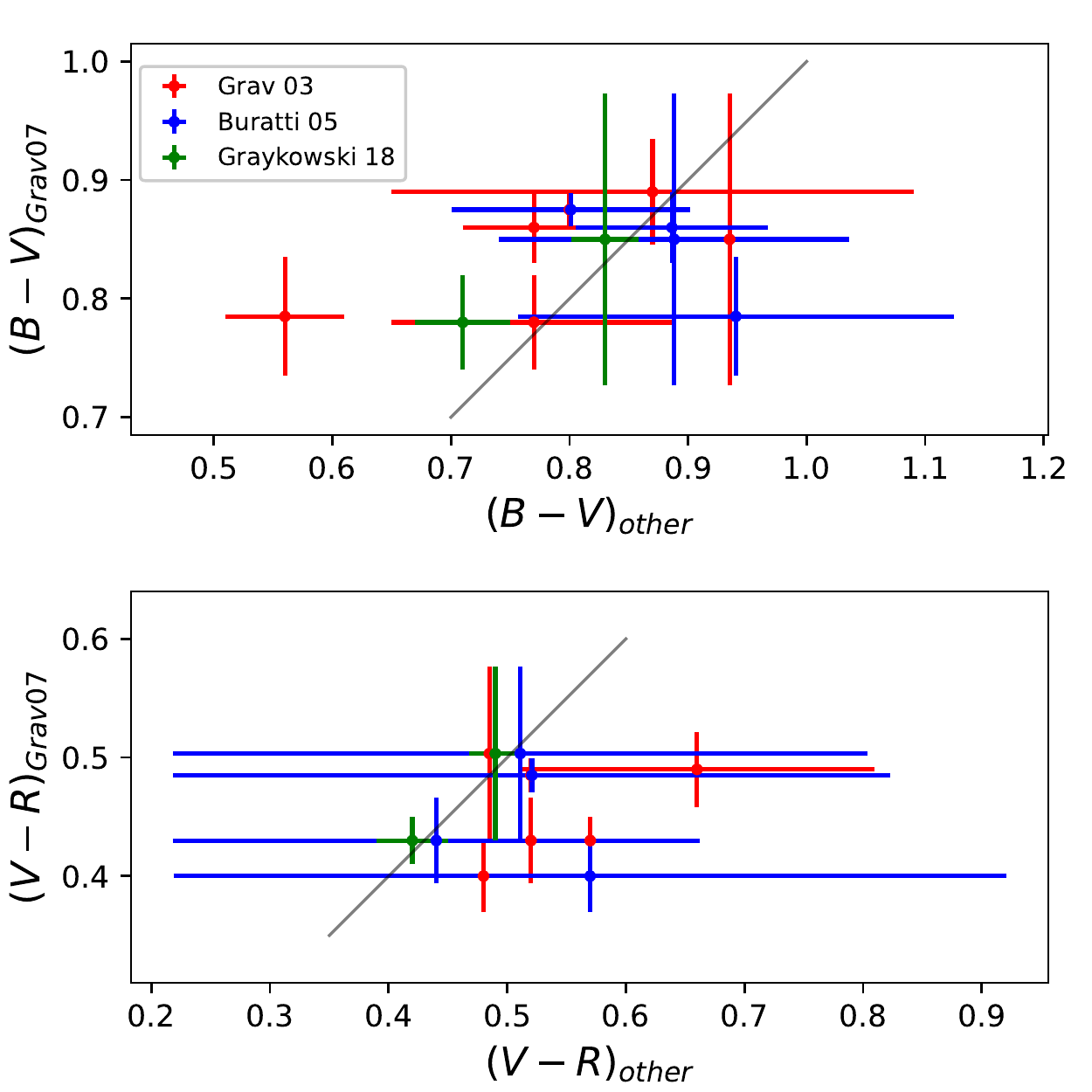}
\caption{Same as Figure \ref{fig:color_comparison} but comparing data from \cite{2007Icar..191..267G} (which contains the greatest number of SIrrs colors in past works) against \cite{2003Icar..166...33G} (in red), \cite{2005Icar..175..490B} (in blue) and \cite{2018AJ....155..184G} (in green).}
\label{fig:color_comp_other}
\end{figure}

The main problem with our measurements is the ``time blocks'' system we choose to observe in different bands (see section \ref{sec:data}), with each of the stacked measurements on each band made in different parts of the SIrrs rotational phases. 
We averaged all observations on each filter to get a proxy of the mean magnitude through the SIrr' rotations. Sadly, we do not have any guarantee that we did not miss any important part of our SIrrs' phase curve. \cite{2019Icar..322...80D} show the lightcurves of many SIrrs at different solar phase angles, showing that these bodies can vary their brightness as much as half a magnitude at low solar phase angle. Although we observed at solar phase angles of $\sim$0.5$^{\circ}$, much lower than in \cite{2019Icar..322...80D}, the variations showed in that work at low phase angles are similar to other brightness variations observed in ground facilities, as seen in \cite{2004ApJ...610L..57B} and \cite{2005Icar..175..490B} for Phoebe and Siarnaq\footnote{In \cite{2005Icar..175..490B} where ``S/2000 S3'' is referred to as Tarvos instead of Siarnaq} respectively. In \cite{2020AJ....159..148P} we explore the color deviations derived from measuring them at different faces of minor bodies' rotations. In the worst case scenario (and assuming constant colors through the SIrrs rotations), that variation is the possible error for the color if observed at opposite peaks of the rotational curve. In our case we can not have observations exactly at opposite peaks since we stack observations in time ranges of a couple of hours, which ``smooths out'' magnitudes variations. Besides of that, for most of the detected SIrrs we average four stacked measurements in $r$ and $i$ bands that are likely to span different parts of the SIrrs, which should also decrease the color error from the ``worst case scenario'' of $\sim$0.5 to our error, which is around 0.2 or 0.3, as seen in Figure \ref{fig:color_comparison}, where although the difference between observations is high, the ``error zero'' line (in black) pass through most of the error ranges. The situation is a bit worse for the $g-r$ colors where the difference between observations gets higher, which can be explained by the less number of ``time blocks'' in $g$ that makes more likely to miss important part of the rotational phase in that band, making our magnitudes more likely to deviate from their actual mean. A good example of this is Phoebe, for which we have stacked colors (from table \ref{tab:colors_stack}) and ``instantaneous'' colors (from table \ref{tab:colors_short}), where it is possible to see that the stacked colors are barely within the errors of the instantaneous measurements (which should be considered as the ``actual'' colors of those sources). Another source of error can be color variations. In the analysis above we assumed that the colors are constant, but in table \ref{tab:colors_short} we see that they can vary as much as 0.36 (in the case of Albiorix in $g-r$). This color variations could also explain the dispersion in measurements in Figure \ref{fig:color_comparison}.
These sources of uncertainty do not only affect our stacked measurements, but also previous surveys, as is shown in Figure \ref{fig:color_comp_other}, where we compare past works against \cite{2007Icar..191..267G} (we show $BVR$ colors for \citealt{2005Icar..175..490B} using the equations from \citealt{2005AJ....130..873J}). A better observational plan to observe SIrrs (or any other Irr) would be to measure consecutively in each filter (for example, in a $ggri$ order) to get observations spanning the same rotational phases. Stacking those observations would increase the signal-to-noise ratio (reducing the errors from observing these faint sources with only one exposure) while enabling a better approximation to a ``mean color'' of the Irrs (at least at the observed rotational phases).

\section{Conclusions} \label{sec:conclusions}

We report $r-i$ color measurements for 21 Saturn Irregular satellites (16 of them with $g-r$ colors), the highest number reported in one single survey for these bodies. (\citealt{2018AJ....155..184G} reported colors for 13 SIrrs in $B-R$ with only 5 of them in $B-V$ and $V-R$, while \citealt{2007Icar..191..267G} reported colors for 12 SIrrs in $B-V$ and $V-R$). Colors of four of them were measured for the first time (Skoll, Hyrrokkin, Kari and Loge).

We managed to obtain these faint sources even through Saturn's brightness and only partially photometric skies by carefully accounting and removing the background brightness from the observations before selectively stacking them and improving their signal-to-noise. 

Despite known sources of uncertainty affect our measurements, as discussed in section \ref{sec:discussion}, these results are good enough to analyse the SIrrs' population as a whole and compare it with other populations. There is no significant difference between our results and those of previous works, as shown in Figure \ref{fig:color_comparison}, where the found colors occupy the same region in color-color space. This is also true for Skoll, Hyrrokkin, Kari and Loge, bodies whose colors had not been measured before. Following the terminology used in \cite{2019AJ....158...53T}, these colors can be described as ``neutral'' and ``moderately red'' (with a few near the ``very red'' region). Although Irrs and TNOs are expected to be related (since both would be heirs of the primordial planetesimal disk), there is an obvious lack of ``ultra-red'' bodies among the SIrrs in comparison to the TNOs (as it can be seen by comparing Figure 6 and 8 of \citealt{2018AJ....155..184G} with Figure 2 of \citealt{2018AJ....155...56J} in $BVR$ bands, or, for $gri$ colors, comparing \citealt{2005Icar..175..490B} and our results with TNOs' colors from  \citealt{2017AJ....154..101P}, \citealt{2018PASJ...70S..40T} and \citealt{2019AJ....158...53T}). The same is true for other populations that are thought to be related to the TNOs, such as the Irrs of the other giant planets \citep{2018AJ....155..184G} or Jupiter and Neptunian Trojans \citep{2018AJ....155...56J, 2019Icar..321..426L}. \cite{2002AJ....123.1039J} shows how outgassing of volatile materials (due to solar irradiation) within short-period comets produces resurfacing of bluer inner material, arguing that this could explain how bodies originated in the Kuiper Belt would become bluer if scattered closer to the Sun.
While for the cases of JIrrs, SIrrs and JTs can be argued that there are not ultra-red bodies because those that once were ultra-red got bluer due to their proximity to the Sun, as explained in \cite{2002AJ....123.1039J, 2018AJ....155...56J}, similarly to what happens to centaurs, which get bluer as their perihelion decreases \citep{2015AJ....150..201J}, that does not explains neutral and moderately red colors of the other populations, as made clear by \cite{2018AJ....155...56J} for the case of the Trojans. Another explanation is proposed in \cite{2020AJ....160...46N}, where is argued that current TNOs colors can be explained if there were no ultra-red material in the primordial planetesimal disk at heliocentric distances below 30 or 40 AU, indicating that current TNOs could have received their moderately red populations from scattered inner bodies when the giant planets migrated and solving why there are no ultra-red bodies among Irrs and Trojans (because there were never ultra-red bodies in their parent populations, at $<30$AU from the Sun). And yet, if the latter were true, we would expect to find more ``neutral'' colors among the TNOs. This lack of neutral bodies for the TNOs could be explained by the continuous thermal effect of the Sun (as explained in \citealt{2002AJ....123.1039J}) and by an evolutionary history marked by collisions \citep{2003AJ....126..398N, 2013Icar..223..775B, 2010AJ....139..994B, 2021Icar..35814184W, 2021PSJ.....2..158A} that could have blued originally moderately red Irrs. In \cite{2021PSJ.....2..158A} is reported a high number of small SIrrs (possibly consequence of a recent impact) evidencing a history richer in impacts than the JIrrs', which could explain the high number of ``neutral'' bodies among the SIrrs, population that appears as blue as the JIrrs population (as seen in \citealt{2018AJ....155..184G}) despite the latter is much closer to the Sun and its radiation.

Finally, we remark the importance of our stacking method which allowed us to get low photometric errors in non-photometric conditions and near a bright source (such as Saturn). By carefully removing the background brightness and selecting images where our targets are not contaminated with stars or any other apparent feature, without having to scale fluxes or matching PSF kernels between images, nor having to point at typical standard stars, but using those in the observed fields with low photometric error and low variability. This should be taken into account for future surveys of similar populations, in particular deep stares around crowded fields as is the plan for part of the Vera Rubin Survey. The same holds for the filter cadence when planning stacking observations.

\subsection*{Acknowledgments}
J.P. acknowledges the support from ANID Chile through (CONICYT-PFCHA / Doctorado-Nacional / 2017-21171752).
J.P. and C.F. acknowledge support from the BASAL Centro de Astrof\'isica y Tecnolog\'ias Afines (CATA) PFB-06/2007, ANID BASAL project AFB-170002 and ANID BASAL project FB210003.
J.P. and C.F. acknowledge support from the Ministry of Economy, Development, and Tourism’s Millennium Science Initiative through grant IC120009, awarded to The Millennium Institute of Astrophysics (MAS).
This project used data obtained with the Dark Energy Camera (DECam), which was constructed by the Dark Energy Survey (DES) collaboration. Funding for the DES Projects has been provided by the US Department of Energy, the US National Science Foundation, the Ministry of Science and Education of Spain, the Science and Technology Facilities Council of the United Kingdom, the Higher Education Funding Council for England, the National Center for Supercomputing Applications at the University of Illinois at Urbana-Champaign, the Kavli Institute for Cosmological Physics at the University of Chicago, Center for Cosmology and Astro-Particle Physics at the Ohio State University, the Mitchell Institute for Fundamental Physics and Astronomy at Texas A\&M University, Financiadora de Estudos e Projetos, Fundação Carlos Chagas Filho de Amparo à Pesquisa do Estado do Rio de Janeiro, Conselho Nacional de Desenvolvimento Científico e Tecnológico and the Ministério da Ciência, Tecnologia e Inovação, the Deutsche Forschungsgemeinschaft and the Collaborating Institutions in the Dark Energy Survey.
The Collaborating Institutions are Argonne National Laboratory, the University of California at Santa Cruz, the University of Cambridge, Centro de Investigaciones Enérgeticas, Medioambientales y Tecnológicas–Madrid, the University of Chicago, University College London, the DES-Brazil Consortium, the University of Edinburgh, the Eidgenössische Technische Hochschule (ETH) Zürich, Fermi National Accelerator Laboratory, the University of Illinois at Urbana-Champaign, the Institut de Ciències de l'Espai (IEEC/CSIC), the Institut de Física d'Altes Energies, Lawrence Berkeley National Laboratory, the Ludwig-Maximilians Universität München and the associated Excellence Cluster Universe, the University of Michigan, NSF's NOIRLab, the University of Nottingham, the Ohio State University, the OzDES Membership Consortium, the University of Pennsylvania, the University of Portsmouth, SLAC National Accelerator Laboratory, Stanford University, the University of Sussex, and Texas A\&M University.
Based on observations at Cerro Tololo Inter-American Observatory, NSF's NOIRLab (NOIRLab Prop. ID 2019A-0915; PI: J. Pe\~na), which is managed by the Association of Universities for Research in Astronomy (AURA) under a cooperative agreement with the National Science Foundation.

%

\vspace{5mm}
\facility{Blanco (DECam)}


\software{astropy \citep{2013A&A...558A..33A,2018AJ....156..123A},
          astroquery \citep{2019AJ....157...98G},  
          Source Extractor \citep{1996A&AS..117..393B},
          photutils \citep{larry_bradley_2020_4049061}
          }

\appendix

\section{Satellite Magnitudes} \label{appendix:colors}

We measured SIrrs magnitudes for every night they were detected, in each band. Note that in tables \ref{tab:mags_all} and \ref{tab:colors_stack} are satellites with no observations in $g$. In those cases we solve equation \ref{eq:phot} without the term associated to $g-r$ to transform their fluxes to magnitudes.

\startlongtable
\begin{deluxetable}{lc|C C C|C C C}
\tablecaption{Satellites' nightly magnitudes
\label{tab:mags_all}}
\tablecolumns{8}
\tablewidth{0pt}
\tablehead{
\colhead{Name} & \colhead{Night} & 
\colhead{$g$\tablenotemark{a}} & \colhead{$r$\tablenotemark{a}} & \colhead{$i$\tablenotemark{a}} &
\colhead{$N_g$\tablenotemark{b}} & \colhead{$N_r$\tablenotemark{b}} & \colhead{$N_i$\tablenotemark{b}}
}
\startdata
Phoebe & 1 & 16.77 \pm 0.03 & 16.52 \pm 0.03 & 16.47 \pm 0.03 & 70 & 30 & 34 \\
 & 2 &  & 16.45 \pm 0.03 & 16.43 \pm 0.03 & 0 & 34 & 31 \\
 & 3 & 16.76 \pm 0.03 & 16.52 \pm 0.02 & 16.39 \pm 0.02 & 53 & 27 & 35 \\
 & 4 &  & 16.47 \pm 0.03 & 16.42 \pm 0.03 & 0 & 21 & 27 \\
\hline
Siarnaq & 1 &  & 20.47 \pm 0.03 & 20.26 \pm 0.05 & 0 & 23 & 29 \\
 & 2 & 21.01 \pm 0.04 & 20.47 \pm 0.03 & 20.19 \pm 0.02 & 41 & 34 & 35 \\
 & 3 &  & 20.47 \pm 0.03 & 20.26 \pm 0.04 & 0 & 26 & 34 \\
 & 4 & 20.99 \pm 0.03 & 20.48 \pm 0.02 & 20.22 \pm 0.02 & 38 & 18 & 30 \\
\hline
Albiorix & 1 &  & 21.05 \pm 0.04 & 20.76 \pm 0.04 & 0 & 13 & 6 \\
 & 2 & 21.66 \pm 0.05 & 21.13 \pm 0.05 & 20.77 \pm 0.04 & 7 & 34 & 35 \\
 & 3 &  & 21.07 \pm 0.04 & 20.77 \pm 0.04 & 0 & 29 & 18 \\
 & 4 & 21.41 \pm 0.03 & 21.03 \pm 0.04 & 20.73 \pm 0.04 & 25 & 29 & 11 \\
\hline
Ymir & 1 & 22.65 \pm 0.03 & 22.16 \pm 0.04 & 22.09 \pm 0.04 & 49 & 9 & 22 \\
 & 2 &  & 22.34 \pm 0.07 & 21.88 \pm 0.06 & 0 & 14 & 18 \\
 & 3 & 22.85 \pm 0.04 & 22.11 \pm 0.05 & 22.11 \pm 0.04 & 47 & 35 & 12 \\
 & 4 &  & 22.38 \pm 0.05 & 21.91 \pm 0.03 & 0 & 21 & 20 \\
\hline
Tarvos & 1 & 23.37 \pm 0.03 & 22.87 \pm 0.02 & 22.59 \pm 0.02 & 46 & 22 & 21 \\
 & 2 &  & 22.90 \pm 0.03 & 22.57 \pm 0.02 & 0 & 6 & 9 \\
 & 3 & 23.43 \pm 0.04 & 22.81 \pm 0.02 & 22.57 \pm 0.02 & 28 & 22 & 23 \\
 & 4 &  & 22.95 \pm 0.03 & 22.58 \pm 0.03 & 0 & 17 & 25 \\
\hline
Kiviuq & 1 & 23.16 \pm 0.03 & 22.31 \pm 0.03 & 22.12 \pm 0.03 & 50 & 28 & 18 \\
 & 2 &  & 22.70 \pm 0.03 & 22.32 \pm 0.03 & 0 & 10 & 9 \\
 & 3 & 22.85 \pm 0.03 & 22.55 \pm 0.03 & 22.36 \pm 0.03 & 27 & 12 & 17 \\
 & 4 &  & 22.40 \pm 0.03 & 22.03 \pm 0.03 & 0 & 15 & 15 \\
\hline
Paaliaq & 1 &  & 21.55 \pm 0.03 & 21.26 \pm 0.03 & 0 & 22 & 24 \\
 & 4 &  & 21.47 \pm 0.03 & 21.27 \pm 0.02 & 0 & 27 & 9 \\
\hline
Ijiraq & 1 & 23.88 \pm 0.02 & 23.21 \pm 0.03 & 22.96 \pm 0.03 & 59 & 27 & 19 \\
 & 2 &  & 23.14 \pm 0.03 & 23.00 \pm 0.03 & 0 & 28 & 18 \\
\hline
Skathi & 1 &  & 24.26 \pm 0.03 & 24.24 \pm 0.04 & 0 & 18 & 26 \\
 & 2 & 24.54 \pm 0.04 &  & 24.28 \pm 0.02 & 34 & 22 & 35 \\
 & 4 & 24.45 \pm 0.03 &  &  & 35 & 14 & 14 \\
\hline
Bebhionn & 1 &  & 24.68 \pm 0.03 & 24.54 \pm 0.03 & 38 & 31 & 29 \\
 & 2 &  &  & 24.60 \pm 0.03 & 0 & 7 & 16 \\
 & 3 &  & 24.65 \pm 0.03 & 24.11 \pm 0.03 & 37 & 17 & 22 \\
\hline
Erriapus & 1 &  & 23.36 \pm 0.03 & 23.08 \pm 0.04 & 0 & 14 & 19 \\
 & 2 & 23.77 \pm 0.07 &  & 23.60 \pm 0.02 & 47 & 12 & 21 \\
 & 3 &  & 24.33 \pm 0.03 & 23.38 \pm 0.03 & 0 & 20 & 10 \\
 & 4 & 24.00 \pm 0.02 & 23.36 \pm 0.02 & 23.00 \pm 0.02 & 28 & 12 & 8 \\
\hline
Skoll & 1 &  & 25.04 \pm 0.03 &  & 0 & 18 & 15 \\
 & 3 &  & 25.00 \pm 0.03 & 24.66 \pm 0.03 & 0 & 28 & 30 \\
\hline
Hyrrokkin & 1 &  & 23.98 \pm 0.03 & 23.74 \pm 0.04 & 0 & 19 & 22 \\
 & 2 &  &  & 23.91 \pm 0.03 & 37 & 28 & 11 \\
 & 3 &  & 24.04 \pm 0.03 & 23.82 \pm 0.04 & 0 & 10 & 28 \\
 & 4 & 24.58 \pm 0.02 & 24.22 \pm 0.03 & 23.49 \pm 0.03 & 24 & 15 & 5 \\
\hline
Mundilfari & 1 & 24.73 \pm 0.03 & 24.25 \pm 0.03 & 24.48 \pm 0.03 & 42 & 35 & 24 \\
 & 2 &  & 24.33 \pm 0.03 & 24.32 \pm 0.03 & 0 & 33 & 32 \\
 & 3 &  & 24.50 \pm 0.03 & 24.29 \pm 0.03 & 29 & 26 & 22 \\
 & 4 &  &  & 24.23 \pm 0.03 & 0 & 8 & 27 \\
\hline
Narvi & 1 & 24.75 \pm 0.02 & 24.41 \pm 0.03 & 24.10 \pm 0.03 & 42 & 21 & 17 \\
 & 2 &  & 24.56 \pm 0.03 & 24.04 \pm 0.03 & 0 & 15 & 32 \\
 & 3 & 24.70 \pm 0.02 & 24.47 \pm 0.03 & 23.98 \pm 0.03 & 31 & 21 & 26 \\
 & 4 &  & 24.48 \pm 0.03 & 24.14 \pm 0.04 & 0 & 18 & 9 \\
\hline
Suttungr & 1 & 24.67 \pm 0.03 & 24.43 \pm 0.03 & 24.20 \pm 0.03 & 32 & 35 & 31 \\
 & 2 &  &  & 24.04 \pm 0.03 & 0 & 21 & 18 \\
 & 3 & 24.78 \pm 0.03 & 24.22 \pm 0.03 & 24.05 \pm 0.03 & 49 & 28 & 22 \\
 & 4 &  &  & 24.35 \pm 0.04 & 0 & 10 & 8 \\
\hline
Bestla & 1 &  &  & 24.29 \pm 0.04 & 0 & 16 & 34 \\
 & 2 &  &  & 24.22 \pm 0.02 & 31 & 27 & 26 \\
 & 3 &  & 24.64 \pm 0.03 & 24.12 \pm 0.03 & 0 & 29 & 22 \\
 & 4 & 25.36 \pm 0.03 &  & 24.42 \pm 0.02 & 37 & 22 & 29 \\
\hline
Thrymr & 1 &  & 24.20 \pm 0.03 & 24.06 \pm 0.04 & 0 & 14 & 16 \\
 & 2 &  &  & 23.95 \pm 0.03 & 33 & 18 & 27 \\
 & 3 &  & 24.02 \pm 0.03 & 23.82 \pm 0.04 & 0 & 15 & 25 \\
 & 4 & 24.48 \pm 0.02 & 24.01 \pm 0.03 & 23.90 \pm 0.03 & 29 & 30 & 16 \\
\hline
Kari & 1 & 24.95 \pm 0.03 &  &  & 55 & 21 & 24 \\
 & 2 &  & 24.50 \pm 0.03 &  & 0 & 34 & 23 \\
 & 3 &  & 24.80 \pm 0.03 & 24.12 \pm 0.04 & 26 & 33 & 20 \\
\hline
Loge & 2 &  & 24.80 \pm 0.03 &  & 0 & 34 & 19 \\
 & 3 &  &  & 24.66 \pm 0.03 & 26 & 14 & 35 \\
\hline
Fornjot & 1 &  & 25.00 \pm 0.03 &  & 41 & 35 & 31 \\
 & 3 &  & 24.86 \pm 0.03 &  & 26 & 24 & 25 \\
 & 4 &  &  & 24.73 \pm 0.04 & 0 & 18 & 16 \\
\enddata
\tablenotetext{a}{Magnitude calculated from the stacked flux at each night.}
\tablenotetext{b}{Number of stacked images per each band.}
\end{deluxetable}


\bibliography{sample631.bib}{}
\bibliographystyle{aasjournal}
\listofchanges

\end{document}